\documentclass[showpacs,amsmath,amssymb,floatfix,prd,superscriptaddress,footinbib,twocolumn]{revtex4-1}
\usepackage{amssymb,scalerel}
\usepackage{graphics}
\usepackage[T1]{fontenc}
\usepackage[utf8]{inputenc}
\usepackage{epstopdf}
\usepackage{epsfig}
\usepackage{dcolumn}
\usepackage{bm}
\usepackage{mathtools}
\renewcommand{\vec}[1]{\mathbf{#1}}
 
\newcommand{\abs}[1]{\left| #1 \right|} 
\newcommand{\avg}[1]{\left< #1 \right>} 
 
\let\baraccent=\= 
\renewcommand{\=}[1]{\stackrel{#1}{=}} 
 
\usepackage{color}
 \definecolor{blue}{rgb}{0,0,1} 
 \definecolor{sepia}{rgb}{0,0.8,0.2}
 \definecolor{redi}{rgb}{0.5176,0.0078,0.0078}

\begin{document}

\title{Buckling  of 2D  Plasma Crystals with Non-reciprocal Interactions}



\author{A. V. Zampetaki}
\email[]{zampetak@uni-duesseldorf.de}
\affiliation{Max-Planck-Institut f\"{u}r Extraterrestrische Physik, 85741 Garching, Germany}
\affiliation{Institut f\"{u}r Theoretische Physik II, Weiche Materie, Heinrich-Heine-Universit\"{a}t, 40225 D\"{u}sseldorf, Germany}

\author{H. Huang}
\affiliation{College of Science, Donghua University, 201620 Shanghai, People's Republic of China}
\author{C.-R. Du}
\affiliation{College of Science, Donghua University, 201620 Shanghai, People's Republic of China}
\author{H. L\"{o}wen}
\affiliation{Institut f\"{u}r Theoretische Physik II, Weiche Materie, Heinrich-Heine-Universit\"{a}t, 40225 D\"{u}sseldorf, Germany}
\author{A. V. Ivlev}
\email[]{ivlev@mpe.mpg.de}
\affiliation{Max-Planck-Institut f\"{u}r Extraterrestrische Physik, 85741 Garching, Germany}

\date{\today}

\begin{abstract}
Laboratory realizations of 2D  plasma crystals typically involve monodisperse microparticles confined  into horizontal monolayers in radio-frequency (rf) plasma sheaths. This gives rise to the so-called plasma wakes beneath the microparticles. The presence of wakes renders the interactions in such systems non-reciprocal, a fact that can lead to a quite different behaviour from the one expected for their reciprocal counterparts. Here we  examine the buckling of a   hexagonal 2D plasma crystal, occurring as the confinement strength is decreased, taking  explicitly into account the non-reciprocity of the system via a well-established point-particle wake model.
We observe that for a finite wake charge, the monolayer hexagonal  crystal undergoes a transition first to a bilayer hexagonal structure, unrealisable in harmonically confined reciprocal Yukawa systems, and subsequently to a bilayer square structure. Our theoretical results are confirmed by molecular dynamics simulations for experimentally relevant parameters, indicating the potential of their observation in state-of-the-art experiments with 2D complex plasmas.
\end{abstract}

\maketitle
\begin{center}
	\textbf{I. INTRODUCTION}
\end{center}

Monodisperse negatively charged micro-sized particles, levitating in a plasma sheath above a powered radio-frequency (rf)
 electrode, 
  form, under sufficiently strong confinement, a monolayer hexagonal complex plasma crystal \cite{Thomas1994,Thomas1996}. Such crystals offer the possibility to study complex kinetic phenomena in solids \cite{Morfill2009,IvlevBook,Bonitz2010}, such as 
  the crystal melting \cite{Thomas1996,Ivlev2003,Chan2004,Nosenko2009,Ivlev2015,Yurchenko2017,Nosenko2013} and the dynamics of dislocations \cite{Nosenko2007}, on a particle-resolved level. Not only have these studies shed light on the generic behaviour of typical solids,
  but also revealed alternative mechanisms such as the crystal melting induced by a mode coupling instability (MCI) \cite{Ivlev2015, Ivlev2000, Zhdanov2009}. 
  
  The MCI makes its appearance  in monolayer hexagonal complex plasma crystals  due to the existence of wake mediated interactions between the dust particles \cite{Couedel2011,Couedel2010,Qiao2013}. The electric field, essential for the levitation of the particles in the rf discharge,
   perturbs the ionic cloud around them, rendering it highly asymmetric. These
   asymmetric ion clouds, known as "plasma wakes" \cite{Kompaneets2016,Lampe2000,Hou2001,Kompaneets2007}, exert attractive interactions on the negatively charged dust particles, making their  pair interactions  non-reciprocal
   and consequently prohibiting a description of the system in terms of a Hamiltonian \cite{Melzer1999,Ivlev2015b}. Nevertheless, the MCI
   is inhibited when the gas damping is strong enough \cite{Couedel2011}. 
   Then the observation of  further unconventional effects introduced by the plasma wakes, such as distinctive structural changes, can be facilitated.

   The non-Hamiltonian nature of the 2D complex plasma becomes more evident for  sufficiently weak confinements, where the wake-charge interactions become more prominent. In such a case the charged particles  create vertical pairs \cite{Steinberg2001,Nosenko2014}.
   When  the interactions nonreciprocity is large  enough the particles constituting these pairs can  jointly  self-propel as a doublet, providing a remarkable example of emerging activity in complex plasma systems \cite{Bartnick2016}.
   
    Regarding the complex plasma crystal, the vertical pairs of charged particles trigger, for a weak confinement, the formation of vertically aligned  hexagonal layers \cite{Schweigert1996,Melzer1996}. In individual cases of neither very weak nor very strong confinement also the formation of multilayer structures has been observed \cite{Teng2003,IvlevBook}, but so far a systematic experimental study of the structural transitions in the system  is missing. 
  
  Meanwhile several theoretical and numerical studies have systematically examined the instability of the hexagonal monolayer Yukawa crystal  in the context of complex plasma  \cite{Totsuji1997,Qiao2005,Klochkov2007, Klumov2008,Donko2008, Hartmann2009,Pan2019}, classical Wigner crystals \cite{Schweigert1999, Travenek2015,Mazars2008,Samaj2012, Antlanger2016} or charged colloids \cite{Messina2003, Oguz2009a, Oguz2009b, Oguz2012, Peng2010, Grandner2008, Grandner2010,Ramiro2006, Ramiro2009,Satapathy2009,Eshraghi2018}. It has been shown that  under harmonic confinement, for a decreasing confinement strength or increasing density,  the hexagonal monolayer is expected to buckle first to  a hexagonal triple layer crystal and subsequently to a bilayer square \cite{IvlevBook}. For even weaker confinement or higher densities, multi-layer structures are expected to prevail, found occasionally also in complex plasma experiments \cite{Teng2003,IvlevBook}. Although the stability of the bilayer square crystal is prominent in all the aforementioned theoretical and numerical studies, it remains still elusive in 2D complex plasma experiments. 
   A possible reason could be that these theoretical  predictions are questionable in the context of 2D  plasma crystals, since they do not take into account the inherent non-Hamiltonian nature of complex plasma caused by the plasma wakes \cite{TsytovichBook}.
   
     In this paper we systematically study the buckling of 2D  plasma crystals following the  structural
     instability of the hexagonal monolayer crystal. In our theoretical and numerical  investigation we explicitly
     take into account the non-reciprocal character of pair interactions in the system, 
     employing a simplified but successful model of wakes as point-like positive charges located below the dust particles \cite{Melzer1996,Ivlev2001,Yaroshenko2005,Couedel2011}.  Within this model we show that for decreasing confinement strength  the hexagonal monolayer gives its place to a hexagonal triple or bilayer crystalline structure, depending on the value of the effective wake charge. Moreover, we identify a large stability region of the bilayer square structure, along the lines of the theoretical predictions for reciprocal Yukawa interactions \cite{IvlevBook, Totsuji1997, Travenek2015, Oguz2012}.
     A part of this region overlaps with the stability region of the bilayer (or triple) layer hexagonal crystal, marking a regime of bistability which proves to be  a source of hysteresis. Our detailed phase diagram for experimentally relevant values of the parameters, not only extends the existing theoretical results to the case of non-reciprocal interactions but also allows for estimating the conditions under which each of the investigated structures can be found, facilitating their observation in experiments of 2D complex plasma crystals.

 \begin{center}
 	{\textbf{II. THE ONSET OF THE MONOLAYER INSTABILITY}}
 \end{center}
 \begin{center}
	{\textbf{A. The model}}
\end{center}
Ignoring the thermal agitation, the equations of motion  (EOM) for  the dust particles in a 2D mono-disperse complex plasma  read

\begin{equation}
m\ddot{\vec{r}}_i+m\nu\dot{\vec{r}}_i=\sum_{j\neq i} \vec{F}_{int}\left(\vec{r}_i-\vec{r}_j\right)+\vec{F}_{ext}\left(\vec{r}_i\right), \label{eom1}
\end{equation}
where $\vec{r}_i$ is the position of the particle $i$, $m$ is the particle mass and $\nu$ is the damping rate originating from the gas friction. 
Each particle $i$ is subjected to two different kinds of forces, namely 
the interaction  force $\vec{F}_{int}\left(\vec{r}_i-\vec{r}_j\right)$, exerted by any other particle $j$,  and the force of the external confinement 
$\vec{F}_{ext}\left(\vec{r}_i\right)$.

The external force results from the assumed parabolic confinement of the particles in the vertical direction $z$ and therefore reads
\begin{equation}
 \vec{F}_{ext}\left(\vec{r}\right)=-m \Omega_{con}^2 z \vec{n}_z, \label{extf1}
\end{equation}
where it is implied that $\vec{r}=x\vec{n}_x+y\vec{n}_y+z\vec{n}_z$ and $\Omega_{con}$ stands for the eigenfrequency of the confining potential well.

\begin{figure}[htbp]
	\begin{center}
		\includegraphics[width=0.99\columnwidth]{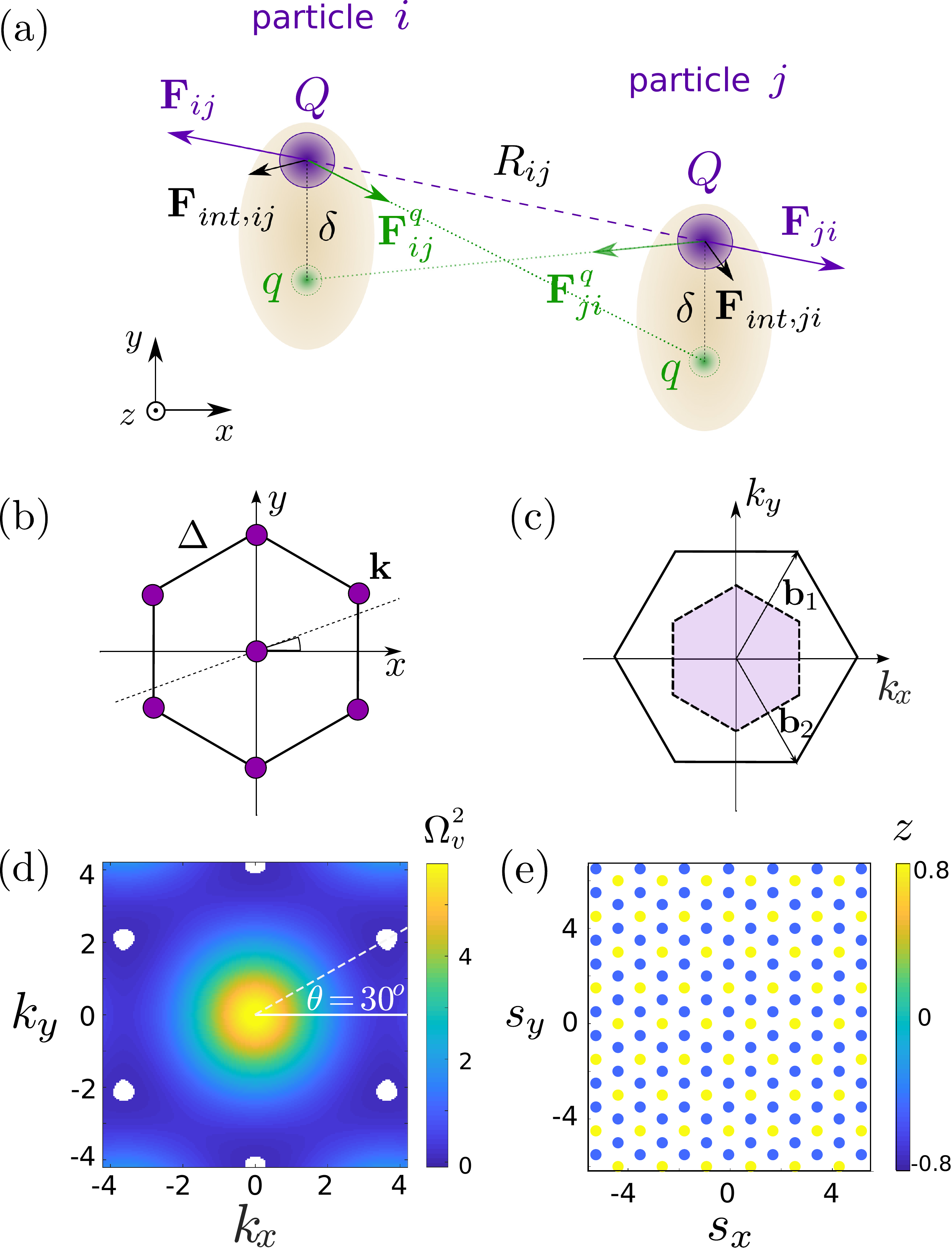}
	\end{center}
	\caption{\label{thr1}  (a) Schematic illustration of the interaction forces $\vec{F}_{int}\left(\vec{r}_i-\vec{r}_j\right)$ between the particles $i$ and $j$. (b) Sketch of the elementary hexagonal lattice cell of the monolayer configuration with the frame of reference.
		Here $\Delta$ denotes the lattice constant and the angle $\theta$ is the angle between the wave-vector  $\vec{k}$ and the $x$ axis.
		(c) Sketch of the reciprocal lattice in $\vec{k}$-space with basis vectors $\vec{b}_{1,2}=2\pi\Delta^{-1}\left(1/\sqrt{3},\pm 1\right)$. 
		The shaded region corresponds to the first Brillouin zone  with boundaries $\left|\vec{k}\right|=2\pi\Delta^{-1}/\sqrt{3}$ for $\theta=0 ^o$ and 
		$\left|\vec{k}\right|=4\pi\Delta^{-1}/3$ for $\theta=30^o$.
		(d) Colour plot of  the squared eigenfrequency of the out-of-plane mode $\Omega_v^2$, pointing to the 
		onset of the structural instability for $\tilde{q}=0.2$, $\tilde{\Omega}_{con}=2.48$, $\tilde{\delta}=0.3$ and $\kappa=1$.
		The white spots indicate where $\Omega_v^2<0$ and
		the structural instability sets in (near $\theta=30^o$
		and $\left|\vec{k}^{(cr)}\right|=4\pi/3$). (e) Colour plot of the $z$ value of each particle for  the
		real part  of the first unstable eigenmode $\vec{d}_{soft}$ as discussed in the text. The particles' positions
		in the $x$-$y$ plane, $s_{x,j}$ and $s_{y,j}$, are depicted, with the colour indicating the position of each particle in the $z$ direction.}
\end{figure}

The interaction force is much more involved since, except from the direct reciprocal interaction between the dust particles $i$ and $j$, we should take also into account the effect of non-reciprocal interactions, stemming from the asymmetry of their surrounding ionic clouds. For the description of the latter
non-reciprocal interactions we employ a simple model, used so far successfully in literature \cite{Melzer1996,Ivlev2001,Yaroshenko2005,Couedel2011} in various situations. Within this model the focusing of ions downstream of the dust particles with a negative charge $Q$, leads to the formation of plasma wakes which can be  approximated by point-like positive charges $q$, located directly below each particle at a fixed distance $\delta$ (Fig. \ref{thr1} (a)).

In this picture the total interaction force $\vec{F}_{int}\left(\vec{R}_{ij}\right)$ exerted on particle $i$ by particle $j$ can be written as a sum of the direct reciprocal interaction between the particles $i$ and $j$ , $\vec{F}_{ij}$ ($\vec{F}_{ij}=-\vec{F}_{ji}$), and the  non-reciprocal interaction between  the particle $i$ and the wake of particle $j$, $\vec{F}_{ij}^q$ ($\vec{F}_{ij}^q\neq-\vec{F}_{ji}^q$), i.e.
\begin{equation}
\vec{F}_{int}\left(\vec{R}_{ij}\right)=Q^2f_Y\left(\vec{R}_{ij}\right)\frac{\vec{R}_{ij}}{{R}_{ij}}+qQf_Y\left(\vec{R}_{ij}^q\right)\frac{\vec{R}_{ij}^q}{{R}_{ij}^q}, \label{intf1}
\end{equation}
(in Gaussian units).

Here we have assumed that both the reciprocal and the non-reciprocal forces are of the Debye-H\"{u}ckel (Yukawa) form 
\begin{equation}
f_Y(R)=\frac{\exp\left(-{R}/{\lambda} \right)}{R^2}\left(1+\frac{R}{\lambda}\right),\label{Yuka}
\end{equation}
with $\lambda$ denoting the effective screening length,
and we have introduced the notation $\vec{R}_{ij}=\vec{r}_i-\vec{r}_j$ and
$\vec{R}_{ij}^q=\vec{r}_i-\vec{r}_j+\delta\vec{n}_z$.
\begin{center}
	{\textbf{B. The emergence of the structural instability}}
\end{center}
The complex plasma system described by Eq. (\ref{eom1}) in view of Eqs. (\ref{extf1})-(\ref{Yuka}) is known to relax at  a hexagonal monolayer crystal   \cite{Thomas1996,Zhdanov2009} for strong enough parabolic confinement.
The reference frame  we use for the description of this  crystal is shown in Fig. \ref{thr1} (b), where  the equilibrium lattice constant is  denoted by $\Delta$ and the angle between the wave vector $\vec{k}$ and the $x$ axis by $\theta$.
The reciprocal lattice is shown in Fig. \ref{thr1} (c) in which also the first Brillouin zone is indicated. Under these assumptions the particles positions in the lattice are given by $\vec{r}_j^{(0)}= \vec{s}^{(H)}_j \Delta$
with 
\begin{equation}
\vec{s}^{(H)}_j=\left(\frac{\sqrt{3}}{2} m_j\right)\vec{n}_x+\left(\frac{1}{2}m_j+n_j\right)\vec{n}_y\label{sH1}
\end{equation}
and $m_j, ~n_j$ specific integers.
 Hereafter, we will use dimensionless units, measuring distance in units of $\Delta$ (i.e. $\Delta=1$)
and frequency in units of the dust lattice frequency
\begin{equation}
\Omega_{DL}=\sqrt{\frac{Q^2}{m\lambda^3}}.\label{omdl1}
\end{equation}
We also introduce the dimensionless quantities of the effective wake charge $\tilde{q}=\abs{{q}/{Q}}$, the effective wake distance $\tilde{\delta}={\delta}/{\Delta}$, the screening parameter $\kappa={\Delta}/{\lambda}$,
and the normalized confinement frequency $\tilde{\Omega}_{con}={\Omega}_{con}/\Omega_{DL}$.

The stability of the hexagonal monolayer crystal can be investigated on the linear level by assuming a plane wave ansatz for  each particle's  displacement $\vec{d}_j$ from its equilibrium position $\vec{r}_j^{(0)}$  as follows
\begin{equation}
\vec{d}_j\propto \exp\left(-i\omega t +i \vec{k}\cdot \vec{r}_j^{(0)}\right). \label{ansatz1}
\end{equation}

The linearization of the EOM (\ref{eom1}) in terms of $\vec{d}_j$   and the use of the above plane wave ansatz lead, as  shown in \cite{Zhdanov2009,Couedel2011,Ivlev2015},  to the dynamical matrix $\vec{D}$
with eigenvalues $\Omega^2_j=\omega_j\left(\omega_j+i \nu\right)$ where $\omega_j$ denote the system's eigenfrequencies.
 Under the assumption $\nu \ll \omega_j$, the hexagonal equilibrium is stable, if all $\Omega_j^2$
are positive.

As discussed in \cite{Zhdanov2009,Couedel2011,Ivlev2015} two of the  $\Omega_j^2$  in addition to their positive real part, attain a finite imaginary part as the confining frequency $\tilde{\Omega}_{con}$ decreases below approximately $3.5$.
This is an imprint of the mode coupling instability (MCI) of the hexagonal monolayer, for weaker confinement strengths, causing  the crystal to melt. Importantly, both this instability and the induced melting can be suppressed by
increasing the damping rate $\nu$, i.e. increasing the gas pressure \cite{Couedel2011}.

For an even weaker confinement, in addition to the MCI, a structural instability  of the monolayer is expected to set in, causing the formation of a multi-layered structure. In order to gain a deeper understanding into the nature of this structural instability we examine the behaviour of    
the squared frequency of the out-of-plane mode $\Omega_{v}^2$ in the $k$-space. When $\Omega_{v}^2<0$, the out-of-plane eigenvector grows exponentially in time and  the system becomes structurally unstable in the vertical direction.
 Thus at the critical value of the confinement frequency for the \textit{structural instability},  $\tilde{\Omega}_{con}^{(SI)}$, the minimum value of $\Omega_{v}^2$
 (for $\vec{k} \neq 0$) becomes zero. As depicted in  Fig. \ref{thr1} (d)
 $\Omega_{v}^2$ attains its lowest value at  $\theta=30^o$ and a wave vector magnitude
 $\left|\vec{k}^{(cr)}\right|=4\pi/3$. This points to the fact that the structural instability sets in at $\vec{k}_{cr}=\frac{2\pi}{\sqrt{3}}\vec{n}_x+\frac{2\pi}{3} \vec{n}_y$. Indeed for a confinement frequency value slightly below $\tilde{\Omega}_{con}^{(SI)}$ we see  that $\Omega_{v}^2$ crosses zero and becomes subsequently negative along the direction of $\vec{k}_{cr}$ (Fig. \ref{thr1} (d)).

The value of $\vec{k}^{(cr)}$ when complemented with the information of the particular crystalline  configuration (Eq. (\ref{sH1})) leads to 
the identification of the softening mode as $\vec{d}_{soft}(m_j,n_j) \propto \exp\left[i\frac{2\pi}{3}(2m_j+n_j) \right]$. The crystalline structure
resulting from this mode is the one described by $s_{z,j}=A \cos\left[\frac{2\pi}{3}(2m_j+n_j) \right]$ (Fig. \ref{thr1} (e)), which can be identified with the hexagonal bilayer (21) with a doubly occupied bottom layer (see Sec. III A). This will be one of our first candidates for the multi-layer structures resulting from the buckling of the hexagonal monolayer.

 \begin{center}
	{\textbf{III. MONOLAYER BUCKLING: THEORY}}
\end{center}
\begin{figure}[htbp]
	\begin{center}
		\includegraphics[width=0.91\columnwidth]{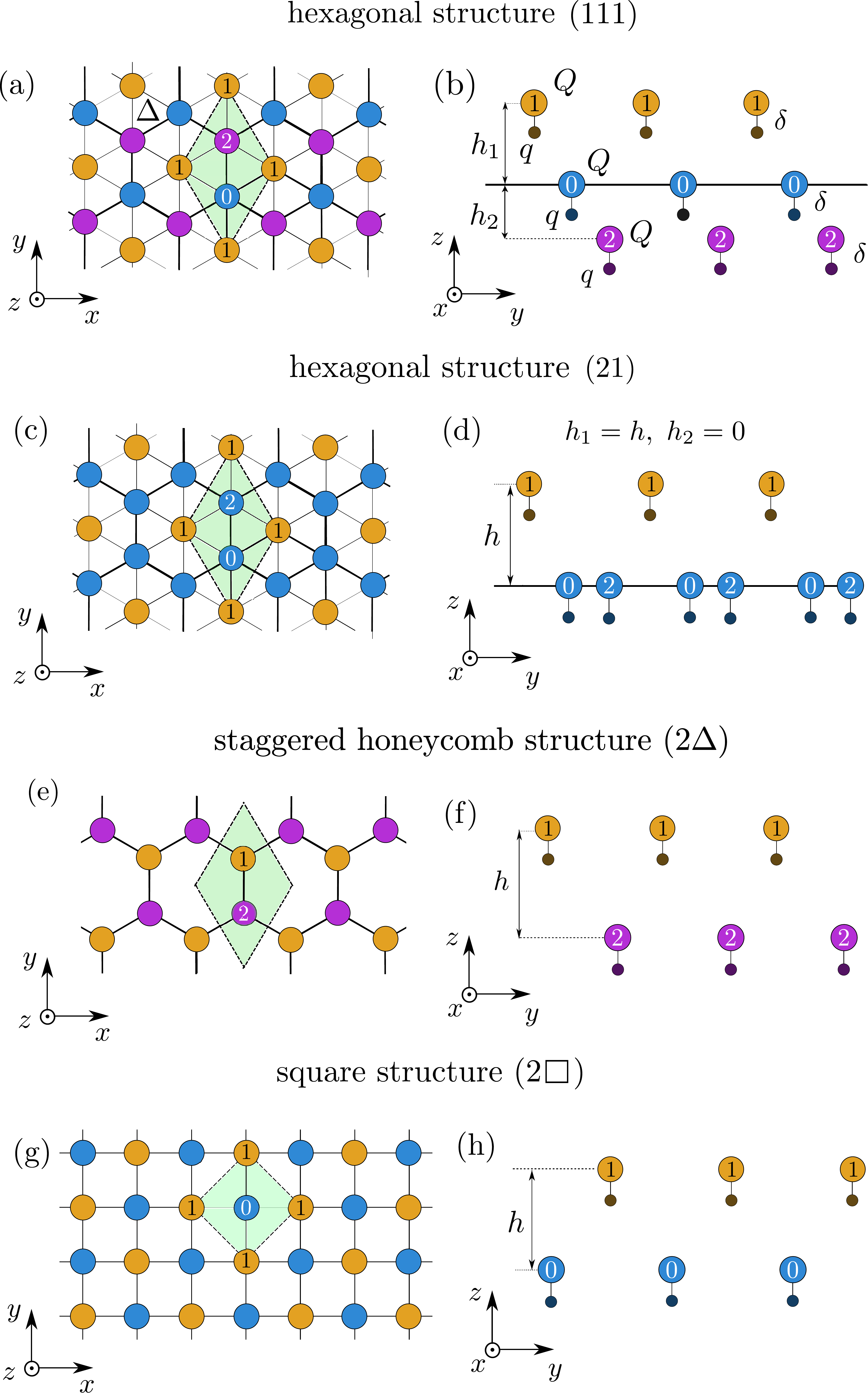}
	\end{center}
	\caption{\label{asy_co1}  Sketch of the 
		different candidate equilibrium structures, investigated in this work: (a),(b) a general hexagonal triple layer structure  $(111)$, (c),(d) a hexagonal bilayer structure with a doubly occupied bottom layer (21), (e),(f) a bilayer staggered honeycomb structure $(2\Delta)$, and (g),(h) a bilayer square structure $(2\Box)$. The left part shows a top view of each structure, while the right part a side view.
		For the triple layer structure, the separation between the middle layer and the top or the bottom layer  is denoted by $h_1$ or $h_2$ respectively. For the bilayer structures the separation between the two layers is denoted simply by $h$. Note that all
		the particles are identical and  the color distinguishes only between particles in the different layers. 
		Also in all the subfigures the numbers $0,1,2$ stand for the corresponding values of $l$ or $w$ as discussed in the text (Eqs. (\ref{leq}), (\ref{weq})) and the shaded green regions  mark the unit cells of the corresponding lattices.}
\end{figure}
Having seen that with decreasing confinement frequency $\tilde{\Omega}_{con}$ the monolayer hexagonal
 crystal undergoes not only a MCI, but also a structural instability at a value $\tilde{\Omega}_{con}^{(SI)}$, we examine in this section the structure of the 2D complex plasma crystal  as $\tilde{\Omega}_{con}$ lowers below $\tilde{\Omega}_{con}^{(SI)}$. We remark here that due to the non-Hamiltonian nature of the system, caused by the presence of wakes, the energy minimization  techniques which are commonly used to tackle theoretically such problems \cite{Totsuji1997,Messina2003,Oguz2009a,Oguz2009b,Oguz2012}, cannot be applied in our case. Therefore, we take the cumbersome route of first solving the force equilibrium equations to identify some of the system's crystalline equilibria, and subsequently using the dynamical matrix to determine the regimes of their stability.

%
 \begin{center}
	{\textbf{A. Candidate structures}}
\end{center}

As discussed above, the first unstable mode of the hexagonal monolayer yields a
bilayer hexagonal structure  (Fig. \ref{thr1} (e)). Having in mind also the results for reciprocal Yukawa systems \cite{Totsuji1997,Oguz2009a,Oguz2012} which predict a bifurcation of the hexagonal monolayer $(1\Delta)$ to a symmetric triple layer hexagonal structure ($3\Delta$), we proceed to the investigation of a more general structure of hexagonal order, i.e. an asymmetric hexagonal triple layer,
termed hereafter as $(111)$  and presented schematically in Fig. \ref{asy_co1} (a),(b). In this configuration and  within the frame of reference of 
Fig. \ref{thr1} (b),
the position of the $j$-th particle in the lattice is given by the vector 
\begin{equation}
\vec{r}_j^{(111)}=\vec{s}_j^{(H)}+s_{z,j}^{(111)} \vec{n}_z \label{hex_r1}
\end{equation}
with  $\vec{s}_j^{(H)}$ defined above in Eq. (\ref{sH1}) and
\begin{equation}
s_{z,j}^{(111)}=\frac{2}{\sqrt{3}}h_l\sin\left(\frac{2\pi}{3}l\right),\label{sz_r1}
\end{equation}
where in order to 
simplify the notation we have introduced
\begin{equation}
l=\text{mod}\left(2m_j+n_j,3\right). \label{leq}
\end{equation}
\\
According to this notation, when $l=0$ the $j$-th particle belongs to the 
middle  layer at a height $h_0=0$ (blue particles in Fig.  \ref{asy_co1} (a),(b)), when  $l=1$ it belongs to  the top layer at a height $h_1$
(yellow particles in Fig. \ref{asy_co1} (a),(b)), whereas when $l=2$ it belongs to  the bottom layer with a height $-h_2$ 
(purple particles in Fig. \ref{asy_co1} (a),(b)). Note that due to the presence of wakes the layers are generally not equidistant, i.e. $h_1\neq h_2$.  

The general triple layer hexagonal configuration (111), depicted in Fig. \ref{asy_co1} (a),(b), can lead, for suitably chosen $h_1$ and 
$h_2$, to different configurations of hexagonal order. The most important for this study are those of an enhanced symmetry. In particular, we obtain the bilayer (21) configuration  with a doubly occupied bottom layer
for  $h_1=h,~h_2=0$ (Fig. \ref{asy_co1} (c),(d)), its reverse bilayer configuration (12) for $h_1=0,~h_2=h$ and the
equidistant triple layer configuration ($3\Delta$) for  $h_1=h_2$. The hexagonal monolayer configuration (1$\Delta$)
is the trivial case $h_1=h_2=0$. Note that all these multi-layer configurations, similarly to the (111) structure 
possess a unit cell consisting of three atoms (Fig. \ref{asy_co1} (a),(c)).

Motivated by the results regarding the buckling of the hexagonal monolayer crystal in colloids \cite{Oguz2009a, Oguz2009b,Oguz2012}, we use here two additional candidate  structures, namely the bilayer staggered honeycomb (2$\Delta$) and the bilayer square (2$\Box$) depicted in Fig. \ref{asy_co1} (e),(f) and Fig. \ref{asy_co1} (g),(h) respectively. Both these structures consist of two layers, separated by a distance $h$ and possess a unit cell consisting  of two atoms. The staggered honeycomb structure can be described by Eqs.(\ref{sH1})-(\ref{leq}),
if we ignore the particles with $l=0$ and keep only those with $l=1,2$. The position of the $j$-th particle in the square bilayer lattice is given in a similar way by
\begin{equation}
\vec{r}_j^{(2\Box)}= m_j\vec{n}_x+n_j\vec{n}_y+s_{z,j}^{(2\Box)} \vec{n}_z \label{sq_r1}
\end{equation}
with $m_j, ~n_j$ specific integers and
\begin{equation}
s_{z,j}^{(2\Box)}=-\frac{1}{2}h_w\cos\left(\pi w\right),\label{sz_s1}
\end{equation}
where  we have introduced
\begin{equation}
w=\text{mod}\left(m_j+n_j,2\right) \label{weq}
\end{equation}
in order to distinguish between the particles belonging to the two different layers.
\begin{center}
	{\textbf{B. Equilibrium structures}}
\end{center}

Evidently all the candidate structures considered here (Fig. \ref{asy_co1}) 
are equilibrated in the $x$-$y$ plane, since due to their symmetry  the $x$ and $y$ components of the total interaction force for each particle vanish. Thus, the only free parameters to be determined are the interlayer separations $h_1,~h_2$ or $h$. Their equilibrium values $h_1^{(0)},~h_2^{(0)}$ or $h^{(0)}$ can be found by demanding a force equilibrium in the $z$ direction.
Since we only care about the interlayer separations our equilibrium condition is that the $z$ component of the net force acting on each layer should be the same with the $z$ component of the net force acting to each other layer. For the triple layer configuration of Fig. \ref{asy_co1} (a),(b) the equilibrium condition thus reads
\begin{eqnarray}
\tilde{F}_{1,0}^{t}+\tilde{F}_{1,2}^t+\tilde{F}_{1,1}^q-\tilde{\Omega}_{con}^2 h_1= \nonumber \\
=\tilde{F}_{0,1}^t+\tilde{F}_{0,2}^t+\tilde{F}_{0,0}^q \nonumber= \nonumber \\
=\tilde{F}_{2,0}^t+\tilde{F}_{2,1}^t+\tilde{F}_{2,2}^q+\tilde{\Omega}_{con}^2 h_2,  \label{equicon}
\end{eqnarray}
where $\tilde{F}_{p,u}^t$ is the total interaction force in the $z$ direction exerted on the layer with $l=p$ from the layer with $l=u$, both from its charges and wakes and  
$\tilde{F}_{p,u}^q$ is the net interaction force in the $z$ direction acting  on the layer with $l=p$ due to the wakes of  the layer with $l=u$. The full expressions for $\tilde{F}_{p,u}^t$ and $\tilde{F}_{p,u}^q$ are provided in Appendix A, but here we remark that both are basically functions  of $h_1$ and $h_2$, so that the solutions of the system (\ref{equicon}) supply us with the equilibrium values $h_1^{(0)},~h_2^{(0)}$. Of course for a bilayer structure such as the ones in Figs. \ref{asy_co1} (e)-(h) the equilibrium condition (\ref{equicon}) reduces to a single equation.

\begin{figure}[htbp]
	\begin{center}
		\includegraphics[width=0.98\columnwidth]{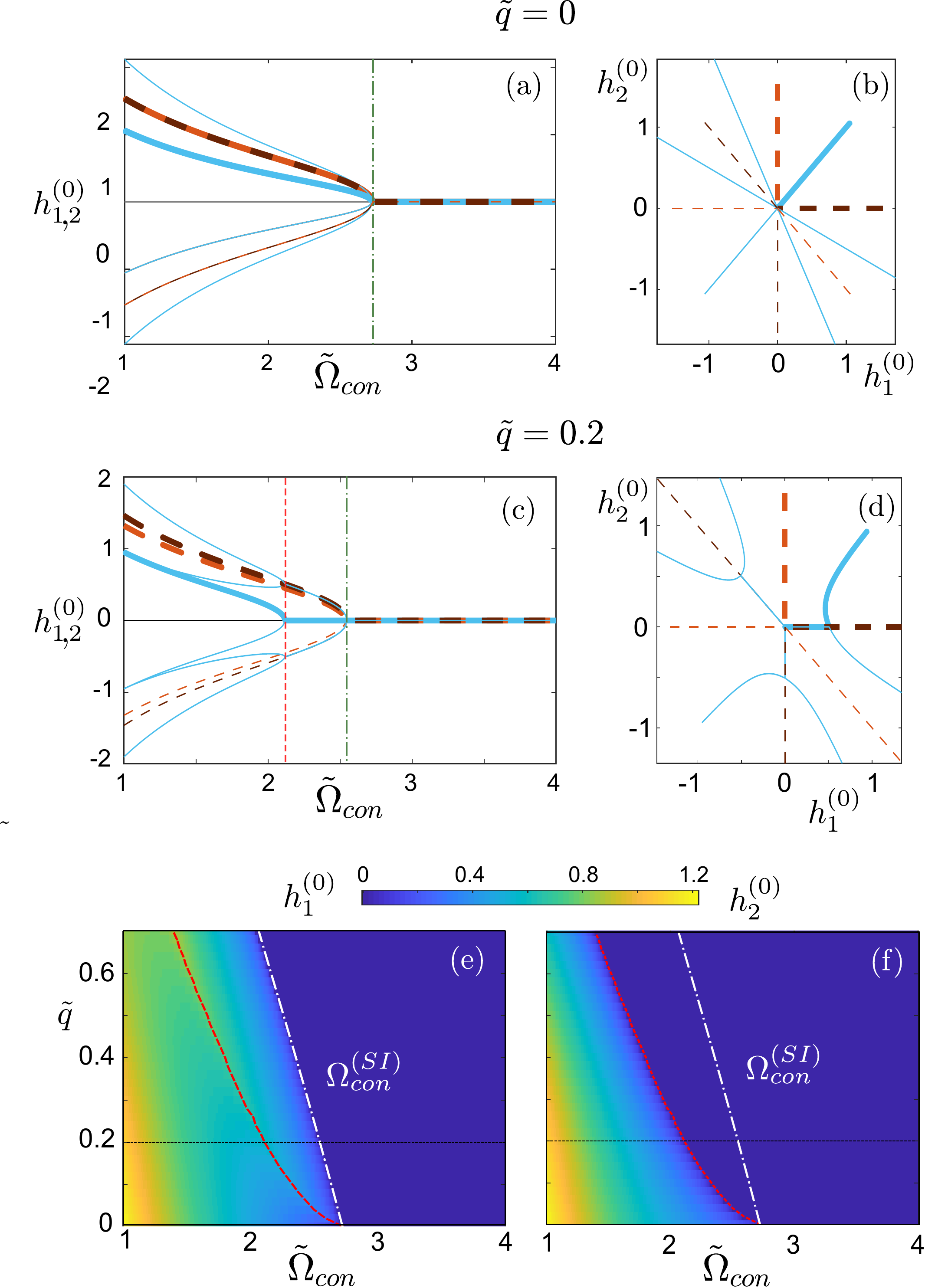}
	\end{center}
	\caption{\label{states1} (a)-(d) Bifurcation diagrams of the hexagonal monolayer configuration in terms of the interlayer distances $h_1^{(0)}$ and $h_2^{(0)}$
		 for two qualitatively different cases  (a),(b) $\tilde{q}=0$ and
		 (c),(d)  $\tilde{q}=0.2$ . In the first column (subfigures (a),(c)) all the possible equilibrium 
		values of $h_1^{(0)}$ (and consequently also $h_2^{(0)}$), are depicted as a function of the confinement frequency $\tilde{\Omega}_{con}$.
		In the second column  (subfigures (b),(d)) all the equilibrium pairs ($h_1^{(0)}$,$h_2^{(0)}$)
		obtained when varying $\tilde{\Omega}_{con}$ are shown, so that we can identify the character of the different equilibria.
		All the light blue lines correspond to triple layer configurations ((111) or ($3\Delta$)), all the brown lines to bilayer ($21$) configurations
		and all the  orange lines to bilayer ($12$) configurations. Note that the different lines of the same colour correspond essentially to the same equilibrium 
		structure and account for all the different permutations of particles $0,1,2$ in Fig.  \ref{asy_co1}  (a),(b). The thicker lines (dashed or not) guide the eye to one of the equivalent structures.
		(e),(f) The values of the equilibrium  separations  (e) $h_1^{(0)}$ and (f) $h_2^{(0)}$   (depicted by colour)
	as a function of $\tilde{\Omega}_{con}$ and $\tilde{q}$. In all the cases we have used $\kappa=1$ and $\tilde{\delta}=0.3$. }
\end{figure}

Using a Newton method with different initial guesses of the values of $h_1,~h_2$ in order to solve numerically Eq. (\ref{equicon}),
we obtain all the possible equilibrium values  $h_1^{(0)}, h_2^{(0)}$ for different confinement frequencies $\tilde{\Omega}_{con}$ and
different values of the effective wake charge $\tilde{q}$. Our results are presented in Fig. \ref{states1} (a)-(d) for the cases of reciprocal interactions ($\tilde{q}=0$, (a),(b)) and non-reciprocal interactions ($\tilde{q}=0.2$, (c),(d)). 

Starting with the case of zero wake charge, $\tilde{q}=0$,
we observe that for high values of $\tilde{\Omega}_{con}$ the only possible equilibrium is that of
the hexagonal monolayer with $h_1^{(0)}=h_2^{(0)}=0$ (Fig. \ref{states1} (a)). Below a critical frequency $\tilde{\Omega}_{con}^{(SI)} \approx 2.7$
the situation changes, and the hexagonal monolayer  bifurcates to three qualitatively different solutions of hexagonal order. Namely, we identify with the help of Fig. \ref{states1} (b)
the bilayer (21) configuration (Fig. \ref{asy_co1} (c),(d)) with $ h_1^{(0)}>0,h_2^{(0)}=0$ (brown bold dashed line), 
the bilayer (12) configuration with $ h_1^{(0)}=0,h_2^{(0)}>0$ (orange bold dashed line) and 
the symmetric triple layer ($3\Delta$) configuration with $ h_1^{(0)}=h_2^{(0)}$ (light blue bold line). Importantly, 
all the three different configurations emerge at the same critical value $\tilde{\Omega}_{con}^{(SI)}$ 
where the monolayer
becomes structurally unstable (Fig. \ref{states1} (a)) and the solution space displays a high degree of symmetry, since the (21) and (12) bilayer solutions are mirror symmetric (Fig. \ref{states1} (a)) and the triple layer solution is equidistant  (Fig. \ref{states1} (b)).

This symmetry breaks for a finite wake charge as shown in Figs. \ref{states1} (c),(d) where  $\tilde{q}=0.2$. The bilayer solutions, although they still arise at the same value of the confining frequency $\tilde{\Omega}_{con}^{(SI)}$, cease to be mirror symmetric, since the interlayer separation $h^{(0)}$ of (21) is larger than that of (12) (Fig. \ref{states1} (c)). This can be seen a consequence of the non-reciprocity of the interactions, considering that in the presence of wakes a double occupation of the top layer (12) exerts to the bottom layer a larger attraction than the one exerted by a doubly occupied bottom layer (21) to the top layer. This wake-induced symmetry breaking affects also 
the triple layer solution.
For a finite wake charge  $\tilde{q}=0.2$, this becomes asymmetric (111) with  $h_1^{(0)}>h_2^{(0)}$ (Fig. \ref{states1} (d)) and  emerges from a bifurcation of the bilayer (21) solution at a lower value of the confinement frequency $\tilde{\Omega}_{con}<\tilde{\Omega}_{con}^{(SI)}$ (Fig. \ref{states1} (c)).

The values of $h_1^{(0)}$ and $h_2^{(0)}$ in the branch of the triple  layer 
(111) solution are presented as a function of the confinement frequency $\tilde{\Omega}_{con}$ and the effective wake charge $\tilde{q}$ in Figs.  \ref{states1} (e) and (f). We can see that the monolayer configuration bifurcates first to the bilayer configuration (21) with $h_2^{(0)}=0$ and subsequently to a triple layer (111) with $h_1^{(0)}>h_2^{(0)}$ for all values of $\tilde{q}$ except for $\tilde{q}=0$ where it bifurcates directly to the (3$\Delta$) structure with $h_1^{(0)}=h_2^{(0)}$. Also the critical  confinement frequencies  for both bifurcations shift to lower  values for increasing $\tilde{q}$.


Regarding the other two candidate structures (Figs. \ref{asy_co1} (e)-(h)), i.e the bilayer honeycomb (2$\Delta$) and the bilayer square structure (2$\Box$) they only lead to a single equilibrium solution, since they possess only one free parameter, the interlayer distance $h$. 

\begin{center}
	{\textbf{C. Stability and phase diagram}}
\end{center}

Having explored the character  of  the different  equilibrium configurations of
Fig. \ref{asy_co1} we next investigate their stability, which is essential for their physical realization.
For this reason we linearize 
our equations of motion (\ref{eom1})  around each of the  equilibria $\{h_1^{(0)},~h_2^{(0)}\}$ or $h^{(0)}$
and assume a plane wave ansatz similarly to Eq. (\ref{ansatz1}) for the displacement $\vec{d}_j$ of the $j$-th particle.
 Following this procedure we construct the corresponding dynamical matrix  
\begin{equation}
\vec{D}=\begin{pmatrix}
\vec{D}_{xx} & \vec{D}_{xy} & \vec{D}_{xz} \\
\vec{D}_{yx} & \vec{D}_{yy} & \vec{D}_{yz}  \\
\vec{D}_{zx} & \vec{D}_{zy} & \vec{D}_{zz}  
\end{pmatrix} \label{Dmat1}.
\end{equation}
In this expression the  $\vec{D}_{uv}$, with  $u,v=x,y,z$ are square 
submatrices of dimension equal to the number of atoms in the unit cell of the examined structure. As an example, for the case of the hexagonal triple layer structure (111) of Fig. \ref{asy_co1} (a),(b), we have that

\begin{equation}
\vec{D}_{uv}=\begin{pmatrix}
D_{uv,00} & D_{uv,01} &  D_{uv,02} \\
D_{uv,10} & D_{uv,11} &  D_{uv,12} \\
D_{uv,20} & D_{uv,21} &  D_{uv,22}
\end{pmatrix} \label{Dmat2}
\end{equation}
with 
\begin{eqnarray}
D_{uv,Ll}&=&a_{uv,L}~\delta_{Ll}+b_{uv,Ll} ~~\text{for $uv \neq zz$} \nonumber \\
D_{zz,Ll}&=&\left(a_{zz,L}+\tilde{\Omega}_{con}^2\right)\delta_{Ll}+b_{zz,Ll} \label{expr1}
\end{eqnarray}
 for $L,l=0,1,2$, $\delta_{ij}$ the Kronecker delta, and  the formulas  for  $a_{uv,L}$, $b_{uv,Ll}$ presented in Appendix B.
Note that in Eq. (\ref{Dmat2}) the $0,1,2$ stand for the different values of $l$
(Eq. (\ref{leq})). A similar expression can be derived also for the other equilibrium  structures explored here, e.g. the square bilayer $(2\Box)$, with a replacement of $l$ with $w$ (Eq. (\ref{weq})). The resulting dynamical matrix $\vec{D}$ has always a dimension $d$ equal to three times the number of particles in the crystal's unit cell.

In order to investigate the stability of each  configuration 
we examine the eigenvalues $\Omega_j^2$ of $\vec{D}$ for the respective equilibria $\{h_1^{(0)},~h_2^{(0)}\}$ or $h^{(0)}$.
In a strict sense stability is provided if and only if  the imaginary part of all eigenfrequencies $\Omega_j$ is equal to zero,
i.e. Im $\Omega_j= 0,~\forall j$. Therefore, a suitable  measure of instability can be provided by
the quantity $S_{int}=\sum_{j=1}^d \left|\text{Im}~\Omega_j \right|$, with it being zero  corresponding to the respective 
configuration being stable and otherwise unstable.  

As we have discussed, however, our system can display apart from structural instability (SI) also mode coupling instability (MCI)
connected to the  appearance of an imaginary part to some of the eigenvalues of $\vec{D}$, $\Omega_j^2$ \cite{Zhdanov2009,Couedel2011}.
This would yield as well Im $\Omega_j \neq 0$ for some $j$, and therefore $S_{int} \neq 0$. In order to  capture only the structural instability we need
to filter out the cases with Re $\Omega_j^2 > 0$ and Im $\Omega_j^2 \neq 0$ from $S_{int}$, i.e. use the measure

\begin{equation}
\tilde{S}_{int}=\sum_{\substack{j\\ \text{Re}~\Omega_j^2<0}}^d \left|\text{Im}~\Omega_j \right|. \label{sint1}
\end{equation}
In the following we judge stability by the values of $\tilde{S}_{int}$. If this is zero for all wave vectors $\vec{k}$ the corresponding configuration
is regarded as structurally stable and otherwise as structurally unstable.

\begin{figure}[htbp]
	\begin{center}
		\includegraphics[width=0.95\columnwidth]{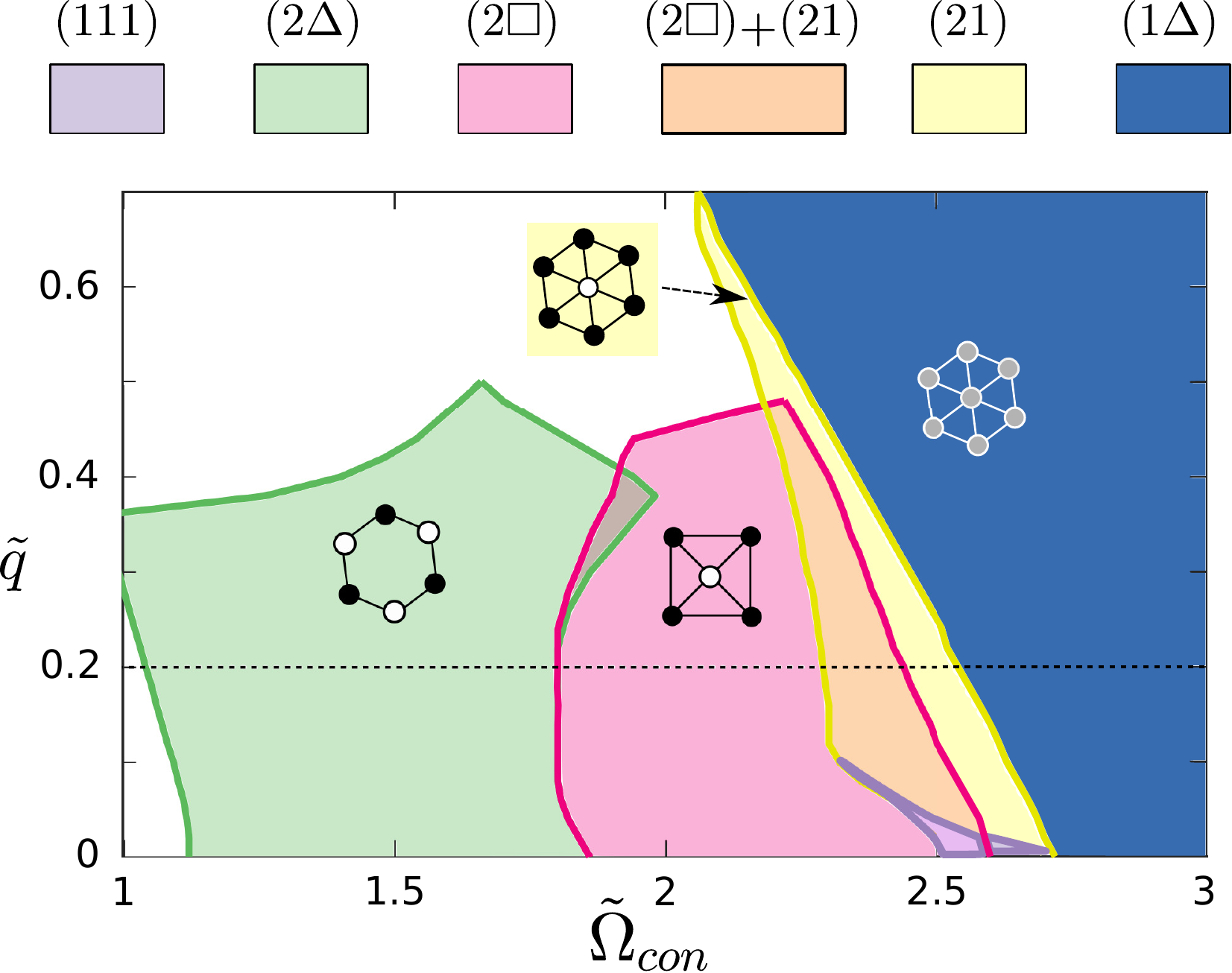}
	\end{center}
	\caption{\label{ph_d1}   The structural phase diagram of a quasi-2D  plasma 
		crystal with non-reciprocal interactions, according to our theoretical findings. The stability regions of the different phases, discussed in the text (Fig. \ref{asy_co1}), are depicted here by different colours.  In this sense the overlap of two different colours
		(e.g. orange region) indicates  a regime of bistability. The control parameters are the confinement frequency $\tilde{\Omega}_{con}$ and the effective wake charge $\tilde{q}$. The other parameters are kept constant at values $\rho=2/\sqrt 3$, $\kappa=1$ and $\tilde{\delta}=0.3$. 
		The horizontal dashed line marks the value of $\tilde{q}=0.2$ used
		in our simulations (Fig. \ref{sim_dz}).
		In the white region none of the examined phases is stable. Note, however, that overall the existence of further stable phases  is not precluded.}
\end{figure}

It turns out that from the structures discussed above only the (12) hexagonal bilayer, with a doubly occupied top layer is always unstable for the 2D complex plasma system. All the other structures explored here (Fig. \ref{asy_co1}) turn to be stable in different parameter regimes. 

The stability regions of each of these structures in terms of the confinement frequency $\tilde{\Omega}_{con}$ and the effective wake charge $\tilde{q}$ are shown in Fig. \ref{ph_d1},
which can be interpreted as the zero-temperature structural phase diagram of our system. Here we observe that for a finite wake charge $\tilde{q}$ with decreasing confinement frequency $\tilde{\Omega}_{con}$ the complex plasma crystal is expected to pass sequentially through the following structures

\begin{equation}
(1\Delta)\rightarrow (21) \rightarrow (2\Box) \rightarrow (2\Delta) \label{stuc_qf}
\end{equation}\\

The reverse sequence is expected for an increasing frequency $\tilde{\Omega}_{con}$ with a hysteresis in the transition  $(2\Box)\rightarrow (21)$, stemming from the evident bistability region between the (21) and $(2\Box)$ structures.
We note here that for typical finite values of $\tilde{q}$ the (21) structure becomes unstable before bifurcating to the hexagonal (111) structure  (Fig, \ref{states1} (c)-(f)). Thus the asymmetric triple layer structure does not appear before the transition to the bilayer  square $(2\Box)$ apart from very low values of $\tilde{q}$ (Fig. \ref{ph_d1})

 Overall the buckling behaviour of the hexagonal monolayer is found to be quite similar to that  of the recipocal system ($\tilde{q}=0$),
where a decreasing frequency triggers the transitions
 
\begin{equation}
(1\Delta)\rightarrow (3\Delta) \rightarrow (2\Box) \rightarrow (2\Delta), \label{stuc_q0}
\end{equation}\\
as known also from \cite{Oguz2009a,Oguz2009b,Oguz2012}.
The major difference is that the hexagonal monolayer (1$\Delta$) gives its place to the bilayer (21) hexagonal structure for $\tilde{q}>0$ instead of the symmetric triple layer hexagonal structure $(3\Delta)$  for $\tilde{q}=0$. However, when these structures become unstable, for lower $\tilde{\Omega}_{con}$ values, a bilayer square $(2\Box)$ is realized for any value of $\tilde{q}$, suggesting that the non-reciprocity of complex-plasma crystals does not prevent the hexagonal-to-square transition known to take place in reciprocal Yukawa systems \cite{IvlevBook, Totsuji1997, Travenek2015, Oguz2012}.

Nevertheless, we note here that neither the bilayer square ($2\Box$)  nor the 
bilayer honeycomb ($2\Delta$) structure are stable for high values of the wake charge $\tilde{q}>0.5$ (Fig. \ref{ph_d1}). In this region we expect that the 
wake-particle attraction will be very large, causing the formation of vertical pairs which at the end will destabilize the 2D complex plasma crystal \cite{Steinberg2001,Nosenko2014}.
 \begin{center}
	{\textbf{IV. MONOLAYER BUCKLING: SIMULATIONS}}
\end{center}
	In order to check the validity of  our theoretical results discussed above we have performed molecular dynamics (MD) simulations  for our point-wake model (Fig. \ref{thr1} (a)). During our simulation 
time we have been changing the confinement strength $\Omega_{con}$ in small steps (for a more detailed description, see Appendix C).
We performed two different simulations: one starting from the hexagonal monolayer (1 $\Delta$) configuration and slowly decreasing the confinement strength 
and another starting from a square bilayer ($2\Box$) configuration and 
slowly increasing $\Omega_{con}$.

\begin{figure}[htbp]
	\begin{center}
		\includegraphics[width=0.99\columnwidth]{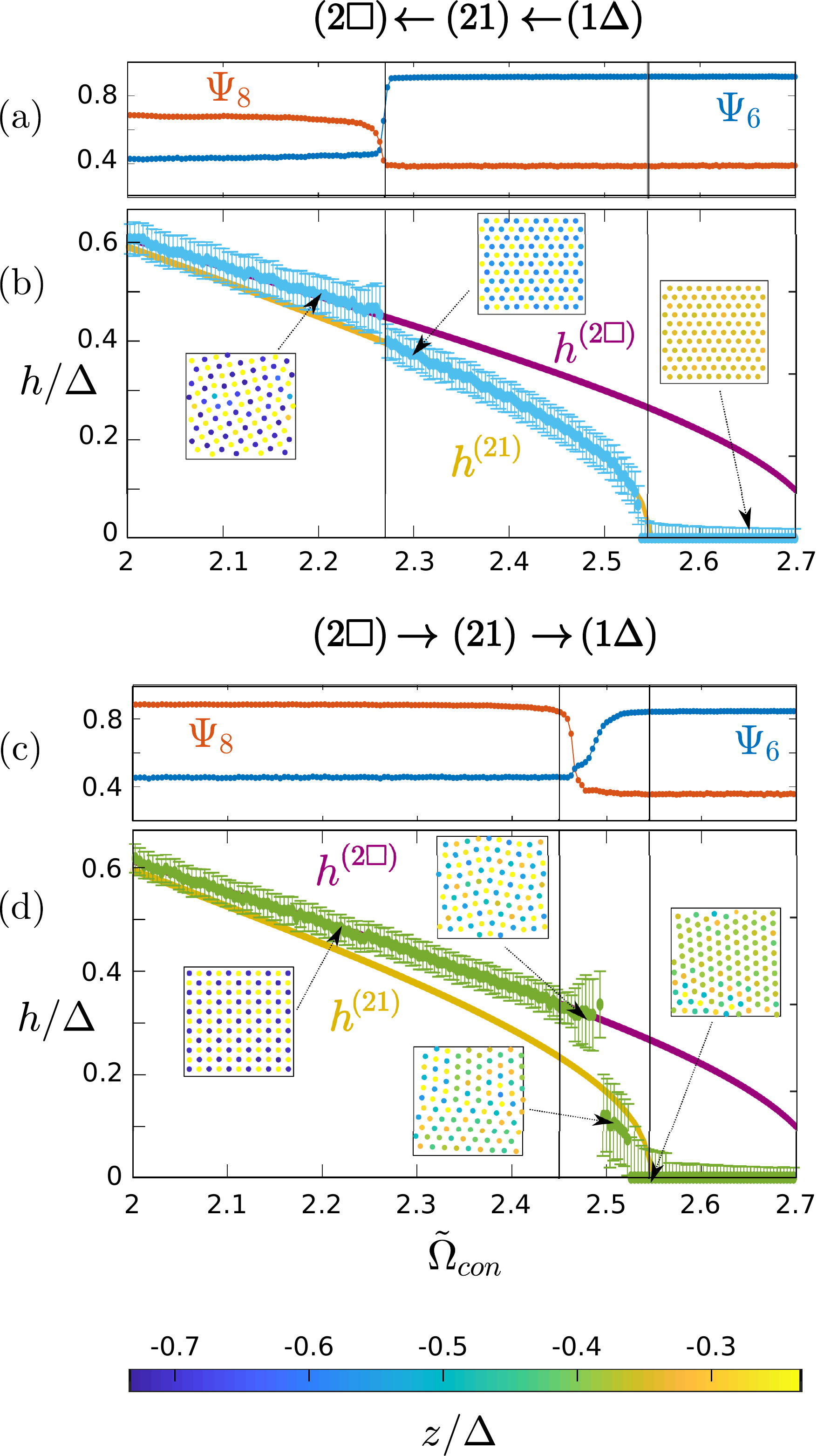}
	\end{center}
	\caption{\label{sim_dz}  (a),(b) MD simulation results  for the buckling of the hexagonal monolayer (1$\Delta$) for experimentally relevant parameters (see Appendix C) as the confinement strength decreases slowly in time. In (a) the evolution of $\Psi_6$ and $\Psi_8$ (Eq. (\ref{Psieq}), used as measures for the hexagonal and the square order respectively,) is depicted as a function of $\tilde{\Omega}_{con}$.  Subfigure (b) shows the evolution of the average interlayer separation $h$ in units of $\Delta$ (light blue markers with errorbars) with decreasing $\tilde{\Omega}_{con}$ (from right to left). For comparison we also show the theoretically expected values for the interlayer separation of the bilayer hexagonal (21) structure $h^{(21)}$ (yellow solid line) and the bilayer square $(2\Box)$ structure $h^{(2\Box)}$ (purple solid line). The square insets are zoomed in snapshots of the system at the indicated by the arrows points. The colour of the particles encodes their vertical positions $z$ in units of $\Delta$. (c),(d) Same as (a),(b) but for the reverse process in which we start by a square bilayer $(2\Box)$ and increase slowly $\tilde{\Omega}_{con}$ (left-to right). The simulation results presented here are for an effective wake charge $\tilde{q}=0.2$, a screening parameter $\kappa=1$ and  a coupling parameter $\Gamma=Q^2/(k_BT \Delta)=1.187\times 10^4$. The vertical lines represent  the theoretically predicted (Fig. \ref{ph_d1}) transition points  for $(1 \Delta)\rightarrow (21)$ and $(21)\rightarrow (2\Box)$ in (a),(b) and for $(2\Box)\rightarrow (21)$ and $(21)\rightarrow (1\Delta)$ in (c),(d).}
\end{figure}

	Our results for the two simulations are presented in Fig. \ref{sim_dz}
  (a),(b) and   Fig. \ref{sim_dz}
  (c),(d) respectively.  Here we focus on the transition from the hexagonal $(1\Delta)$  to the bilayer square $(2\Box)$  structure (decreasing $\tilde{\Omega}_{con}$) and backwards (increasing $\tilde{\Omega}_{con}$). As a measure of the
  degree of hexagonal and square order in the system we use the quantities 
  $\Psi_6$ and $\Psi_8$ respectively, where  the global bond angular order parameter  $\Psi_n$ with $n=1,~2,~3,\ldots$, generally reads
  \begin{equation}
  \Psi_n=\abs{\frac{1}{N}\sum_{j}^N \frac{1}{c_j}~\sum_{k(\text{nn})}^{c_j} \exp\left(in\theta_{j,k}\right)}. \label{Psieq}
  \end{equation}
  In the above expression $N$ is the total number of particles, $c_j$ denotes the coordination number of the particle $j$, $\theta_{j,k}$ is the angle of the bond between the adjacent particles $j$ and $k$ with respect to a fixed reference direction and $k(\text{nn})$ denotes the summation over all particles $k$ which are nearest neighbours of the particle $j$. Within this definition the $\Psi_6$ ($\Psi_8$) is expected to be one for a perfect hexagonal (square)
lattice  and zero if no hexagonal (square) order is present.

As we observe in Fig. \ref{sim_dz} (b), for a confinement frequency $\tilde{\Omega}_{con}$ decreasing slowly with time, the initial hexagonal monolayer (1$\Delta$) bifurcates at a certain value of $\tilde{\Omega}_{con}$ to a clear  bilayer hexagonal structure (21) with a doubly occupied bottom layer. At even lower values of $\tilde{\Omega}_{con}$ the hexagonal structure becomes unstable and several domains of a bilayer square structure ($2\Box$) appear which fill eventually our simulation box. This transition, from the bilayer hexagonal to the bilayer square configuration is accompanied by an abrupt increase of the interlayer separation $h$  (Fig. \ref{sim_dz} (b)), a steep increase of $\Psi_8$ and a subsequent steep decrease of $\Psi_6$  (Fig. \ref{sim_dz} (a)). This behaviour is in line with our theoretical predictions, discussed above (Fig. \ref{ph_d1}, sequence (\ref{stuc_qf})). Even more,  the numerical values for the interlayer separation $h$ and the two transition points are in an excellent agreement with the ones predicted by our theory.

The results for the reverse scenario  (Fig. \ref{sim_dz} (c),(d)), in which we start from a bilayer square configuration ($2\Box$) and increase the confinement frequency $\tilde{\Omega}_{con}$, reveal a clear  hysteretic behaviour, signified by the decrease of the square order ($\Psi_8$) and increase of the hexagonal order  ($\Psi_6$)  at a larger $\tilde{\Omega}_{con}$ value than the one found for the $(21)\rightarrow(2\Box)$ transition (Fig. \ref{sim_dz} (a),(b)). 
The transition point for the loss of square order proves to be well estimated by our theoretical results, in view of which, the cause of the  hysteresis is the bistability  between the $(2\Box)$ and (21) structure in the regarded region (Fig. \ref{ph_d1}). The transition  $(2\Box)\rightarrow (21)$ appears to be overall slower than the $(21) \rightarrow (2\Box)$, especially in the vertical direction where the separation of the particles remains close to the expected one for the bilayer square $(h^{(2\Box)})$ for some interval past the transition (Fig. \ref{sim_dz} (d)). In addition, the structures realized in our simulations after the destabilization of the $(2\Box)$ and before the stabilization of the hexagonal monolayer $(1\Delta)$, i.e. between the two transition lines, are quite disordered, rendering  the identification of  the different layers difficult and resulting in the large deviations in the interlayer separation $h$ (Fig. \ref{sim_dz} (d)).

\begin{figure}[htbp]
\begin{center}
	\includegraphics[width=0.8\columnwidth]{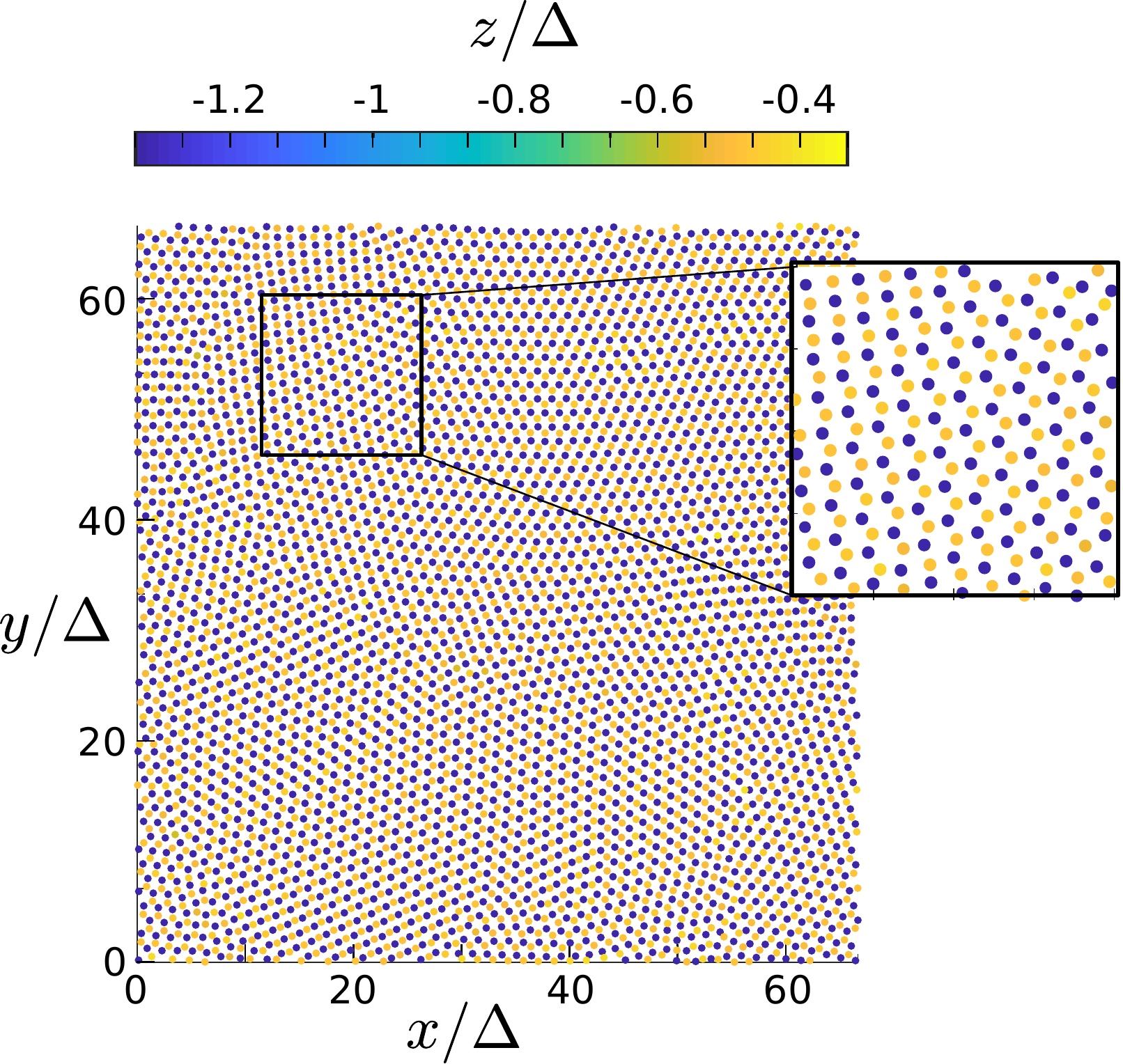}
\end{center}
\caption{\label{ho_si1}    Snapshot of the particles' positions, resulting from MD simulations with slowly decreasing confinement strength, for  the case
	$\tilde{\Omega}_{con}=1.5$. The colour of the particles encodes the values of their vertical positions $z$ in units of $\Delta$. The parameters used read  $\tilde{q}=0.2$, $\kappa=1$ and  $\Gamma=1.187\times 10^4$ (for  more information see  Appendix C).}
\end{figure}

Upon a decrease of its confinement frequency $\tilde{\Omega}_{con}$ below $2$ the system exhibits an even more complex behaviour. Its detailed description, however, goes beyond the scope of the present paper. Here we only mention
that  after the destabilization of the  square bilayer  $(2\Box)$  at a certain  $\tilde{\Omega}_{con}$, we find in our simulations the occurrence of  a bilayer staggered honeycomb $(2\Delta)$ structure (Fig. \ref{ho_si1} ), along the lines of our theoretically calculated phase diagram (Fig. \ref{ph_d1}). Beyond the  destabilization  of the  $(2\Delta)$ structure the system presumably enters a transient regime and at even lower confinement frequencies we expect the successive formation of more layers, similarly to the results of \cite{Teng2003}.

\begin{center}
	{\textbf{V. CONCLUSIONS AND OUTLOOK}}
\end{center}

In this paper we have investigated the buckling of 2D monodisperse complex plasma crystals 
as their vertical confinement weakens. Unlike the Yukawa systems whose buckling has been explored in literature to a large extent \cite{IvlevBook, Totsuji1997, Travenek2015, Oguz2012}, the 2D complex plasma crystals, feature non-reciprocal interactions, due to the presence of the plasma wakes. This fact prohibits their description in terms of a Hamiltonian  \cite{Ivlev2015} and renders the standard minimization techniques unsuitable for the investigation of the structural transitions in the system.

Employing a simple point-particle model for the wakes, we have solved the force equilibrium equations for different lattice structures and determined their stability through the corresponding dynamical matrices, in order to construct the structural phase diagram of the system. Here we have focused on the exploration of the regime close to the instability of the hexagonal monolayer $(1\Delta)$.

For a finite wake  charge, we find that below a critical confinement frequency,
the hexagonal monolayer $(1 \Delta)$ gives its place to a bilayer hexagonal  structure (21) with a doubly occupied bottom layer. This is different from what has been found for  reciprocal Yukawa  systems \cite{Oguz2012}, where instead of the (21) a symmetric triple layer hexagonal $(3\Delta)$ structure is realized, and can be viewed as an imprint of the non-reciprocity, which breaks the system's symmetry. Decreasing further the confinement frequency we observe a discontinuous transition to a bilayer square $(2\Box)$ and subsequently to a bilayer staggered honeycomb $(2\Delta)$ structure, similarly to the reciprocal case.  Our theoretical results are confirmed by molecular dynamics simulations which also show a clear hysteresis of the  $(21) \rightarrow (2\Box)$ transition, owing to the existence of a bistability region of the (21) and $(2\Box)$  structures.

The results presented here for experimentally relevant parameters, confirm the intuition that a bilayer square structure should be realized for 2D monodisperse  complex plasma  crystals at weak enough confinement frequencies. We believe therefore that they can be a useful guide for future experiments aiming at observing  the long-sought bilayer square structures at monodisperse complex plasma systems. The major challenge thereof will be to avoid the  vertical pairing which destabilizes entirely the plasma crystal for high wake charges. Thus, it would be important to determine the experimental  parameter values of the wakes, e.g. by employing existing self-consistent kinetic models for the  wake
 \cite{Kompaneets2007}. 
 
 Finally, we remark that the theoretical procedure employed in this work is quite general and, after adjusted properly, can be applied to explore the structural properties of other non-reciprocal systems, such  as those featuring diffusiophoretic \cite{Sabass2010,Soto2014,Saha2019}, predator-prey \cite{Zhdakin2011,Sengupta2011} or social \cite{Helbing1995,Helbing2000,Gonzalez2006,Nishimoto2013}
 forces.

\appendix
\begin{center}
	{\textbf{APPENDIX A: EXPRESSIONS FOR THE INTERLAYER FORCES}}
\end{center}
Here we provide the full expression for the forces appearing in the equilibrium condition (\ref{equicon}). As a reference example we use the case of a  triple layer hexagonal (111) configuration with interlayer separations $h_1$ and $h_2$, depicted in Fig. \ref{asy_co1} (a),(b). The equilibrium conditions for the other configurations are  obtained following a very similar procedure.

Before we proceed we define the following functions
\begin{eqnarray}
\Lambda(x)&=&\left(x^{-2}+x^{-3}\right)e^{-x} \nonumber \\
\Xi(x)&=&\left(x^{-1}+3x^{-2}+3x^{-3}\right)e^{-x}. \label{lam_psi}
\end{eqnarray}
which will be used in order to express the different forces.

As already mentioned in the text,
 the particles for the (111) configuration can be distinguished into three classes, 
indicative of their position in the unit cell (Fig. \ref{asy_co1}), according to the value
of  $l$ (Eq. (\ref{leq})).
 The value of this index affects the value of the interparticle distances $\tilde{\vec{R}}_{ij}=\left|\vec{r}_i-\vec{r}_j\right|/\Delta$
and consequently the values of $s=\left|\tilde{\vec{R}}_{ij}\right|$ and $s_\delta =\left|\tilde{\vec{R}}_{ij}+\tilde{\delta}\vec{n}_z\right|$ with $\tilde{\delta}=\delta/\Delta$.

For all the cases of (111) configurations we have $s_x=\frac{\sqrt{3}}{2} m$ and $s_y=\frac{1}{2} m+n$, with $m,n$ 
arbitrary integers and additionally  
\begin{eqnarray}
s_{z,l}&=&\frac{2}{\sqrt{3}} \left[ h_l \sin\left(2\frac{\pi}{3}l\right)- h_{r}\sin\left(\frac{2\pi}{3}r\right) \right]\nonumber \\
s^{(\delta)}_{z,l}&=& s_{z,l}+\tilde{\delta} \nonumber \\
s_l&=&\sqrt{s_x^2+s_y^2+s_{z,l}^2} \nonumber \\
s^{(\delta)}_l&=&\sqrt{s_x^2+s_y^2+\left(s^{(\delta)}_{z,l}\right)^2}   \label{exp_s}
\end{eqnarray}
with $h_0=0$, arbitrary $h_1,~h_2$ and \[r=\text{mod}\left[l-(2m+n),3\right].\]

Using these, along with the formula for the interaction forces in our system (Eq. (\ref{intf1})), we can write the expression  for the $z$ component  of the total force $\tilde{F}_{p,u}^t$ exerted on the layer $l=p$ from the layer with $l=u$
as follows
\begin{equation}
\tilde{F}_{p,u}^t=\sum_{\mathclap{\substack{m,n\\ \mod\left[p-(2m+n),3\right]=u}}} ~\left[\Lambda\left(\kappa s_p\right)s_{z,p}-\tilde{q}\Lambda\left(\kappa s^{(\delta)}_p\right)s^{(\delta)}_{z,p}\right]~, \label{fput1}
\end{equation}
where we have also used the $\Omega_{DL}$ unit (Eq. (\ref{omdl1})).
The first sum in this expression corresponds to the interaction between the charges of layer $p$ with  the charges of layer $u$. The second sum, denoted by 
\begin{equation}
\tilde{F}_{p,u}^q=-\sum_{\mathclap{\substack{m,n\\ \mod\left[p-(2m+n),3\right]=u}}} ~\tilde{q}\Lambda\left(\kappa s^{(\delta)}_p\right)s^{(\delta)}_{z,p}~, \label{fpuq1}
\end{equation}
refers to the interactions between the charges of layer $p$ with the wakes of layer $u$. Note that through their dependence on $s_p$, $s_{z,p}$, $s^{(\delta)}_p$
and $s^{(\delta)}_{z,p}$, both  $\tilde{F}_{p,u}^t$ and $\tilde{F}_{p,u}^q$ are functions of  $h_1$ and $h_2$.

For the case of the bilayer square ($2\Box$) the situation is very similar with
the one described above.
The only difference is in the expressions for $s_x=m$, $s_y=n$ and  the use of $w=0$ or $1$
(Eq.( \ref{weq}))  in the place of $l$, which lead to
\begin{eqnarray}
s_{z,w}&=&-\frac{1}{2} \left[ h_w \cos\left(\pi w\right)- h_{b}\cos\left(\pi b\right) \right]\nonumber \\
s^{(\delta)}_{z,l}&=& s_{z,l}+\tilde{\delta} \nonumber \\
s_l&=&\sqrt{s_x^2+s_y^2+s_{z,l}^2} \nonumber \\
s^{(\delta)}_l&=&\sqrt{s_x^2+s_y^2+\left(s^{(\delta)}_{z,l}\right)^2} \label{exp_s2}
\end{eqnarray}
with $h_0=0$, $h_1=h$ and \[b=\text{mod}\left[w-(m+n),2\right].\]

The corresponding expressions for $\tilde{F}_{p,u}^t$ and $\tilde{F}_{p,u}^q$
in this case read
\begin{equation}
\tilde{F}_{p,u}^t=\sum_{\mathclap{\substack{m,n\\ \mod\left[p-(m+n),2\right]=u}}} ~\left[\Lambda\left(\kappa s_p\right)s_{z,p}-\tilde{q}\Lambda\left(\kappa s^{(\delta)}_p\right)s^{(\delta)}_{z,p}\right]~, \label{fput2}
\end{equation}
and 
\begin{equation}
\tilde{F}_{p,u}^q=-\sum_{\mathclap{\substack{m,n\\ \mod\left[p-(m+n),2\right]=u}}} ~\tilde{q}\Lambda\left(\kappa s^{(\delta)}_p\right)s^{(\delta)}_{z,p}. \label{fpuq1}
\end{equation}

\begin{center}
	{\textbf{APPENDIX B: ELEMENTS OF THE DYNAMICAL MATRIX}}
\end{center}
Using the expressions of Appendix A for the (111) configuration
we can write the quantities $a_{pq,L}$, $b_{pq,Ll}$ which construct the dynamical matrix $\vec{D}$, as follows
\begin{eqnarray}
a_{uv,L}&=& \sum_{m,n}   A_{pq,L,m,n}  \nonumber \\
b_{uv,Ll}&=& -\sum_{\mathclap{\substack{m,n\\ \mod\left[L-(2m+n),3\right]=l}}}~A_{pq,L,m,n}\exp\left(-i\vec{k}\cdot\vec{s}\right), \label{ab}
\end{eqnarray}
with $u,v=x,y,z$ and
\\
\begin{widetext}
\begin{eqnarray}
A_{xx,l,m,n}&=& \Xi\left(\kappa s_l\right)\left(s_x/s_l\right)^{2}-\Lambda\left(\kappa s_l\right)-\tilde{q}\left[\Xi\left(\kappa s^{(\delta)}_l\right)\left(s_x/s^{(\delta)}_l\right)^{2}-\Lambda\left(\kappa s^{(\delta)}_l\right)\right]  \nonumber \\
A_{yy,l,m,n}&=&  \Xi\left(\kappa s_l\right)\left(s_y/s_l\right)^{2}-\Lambda\left(\kappa s_l\right)-\tilde{q}\left[\Xi\left(\kappa s^{(\delta)}_l\right)\left(s_y/s^{(\delta)}_l\right)^{2}-\Lambda\left(\kappa s^{(\delta)}_l\right)\right]  \nonumber \\
A_{zz,l,m,n}&=& \Xi\left(\kappa s_l\right)\left(s_{z,l}/s_l\right)^{2}-\Lambda\left(\kappa s_l\right)-\tilde{q}\left[\Xi\left(\kappa s^{(\delta)}_l\right)\left(s^{(\delta)}_{z,l}/s^{(\delta)}_l\right)^{2}-\Lambda\left(\kappa s^{(\delta)}_l\right)\right]  \nonumber \\
A_{xy,l,m,n}&=& A_{yx,l,m,n}=  \Xi\left(\kappa s_l\right)\left(s_xs_y/s_l\right)^{2}-\tilde{q}\Xi\left(\kappa s^{(\delta)}_l\right)\left(s_xs_y/s^{(\delta)}_l\right)^{2}  \nonumber \\
A_{xz,l,m,n}&=& A_{zx,l,m,n}=  \Xi\left(\kappa s_l\right)\left(s_xs_{z,l}/s_l\right)^{2}-\tilde{q}\Xi\left(\kappa s^{(\delta)}_l\right)\left(s_xs^{(\delta)}_{z,l}/s^{(\delta)}_l\right)^{2}  \nonumber \\
A_{yz,l,m,n}&=& A_{zy,l,m,n}=  \Xi\left(\kappa s_l\right)\left(s_ys_{z,l}/s_l\right)^{2}-\tilde{q}\Xi\left(\kappa s^{(\delta)}_l\right)\left(s_ys^{(\delta)}_{z,l}/s^{(\delta)}_l\right)^{2}  \label{Aelms}
\end{eqnarray}
\end{widetext}
In the above expressions we have $l,L=0,1,2$. Replacing $l,L$ with $w,W=0,1$
and using  formulas (\ref{exp_s2}) instead of  (\ref{exp_s}),   we can deduce from Eqs. (\ref{ab}) and (\ref{Aelms}) also the $6 \times 6$ dynamical matrix of the bilayer square $(2\Box)$ configuration.

\begin{center}
{\textbf{APPENDIX C: MOLECULAR DYNAMICS SIMULATIONS}}
\end{center}

Our molecular dynamics (MD) simulations were performed
using LAMMPS in the NVT ensemble \cite{LAMMPS} for the point-particle wake model (Fig. \ref{thr1} (a)). The total number of particles used in our simulations is $N=6480$. In the horizontal direction we apply periodic boundary conditions, while in the vertical direction the 
particles are confined in a parabolic potential, so that the confinement
force  for the particle $i$ reads 

\begin{equation}
\vec{F}_{con,i}= -C(z_i-z_0)\vec{n}_z, \label{MDfc}
\end{equation}
with $C=m\Omega_{con}^2$ the confinement strength and $z_0=0$ the equilibrium point of the confining potential.
The particles' positions are initially chosen so that they form a monolayer hexagonal configuration  in a $24.3~mm \times 20.786~mm$ simulation box. The equations of motion for our simulations, which take into account the damping from the surrounding neutral gas and the 
Brownian motion of the particles due to the finite temperature $T$, read

\begin{equation}
m\ddot{\vec{r}}_i+m\nu\dot{\vec{r}}_i=\sum_{j\neq i} \vec{F}_{int,ij}+\vec{F}_{con,i}+\vec{L}_i, \label{MDeom1}
\end{equation} 
where $\vec{r}_i$ is the position of particle $i$, $m$ the particle mass, $\nu$ the damping  rate and $\vec{F}_{int,ij}$ the non-reciprocal force exerted on particle $i$ from particle $j$ and its wake (Eq. (\ref{intf1})). The 
Langevin force  $\vec{L}_i$ is defined as

\begin{equation}
\avg{\vec{L}_i(t)}=0,~~\avg{\vec{L}_i(t)\vec{L}_j(t+\tau)}=2\nu m k_B T\delta_{ij} \delta(\tau) \vec{I},
\end{equation}
 with $T$ the temperature of the heat bath and $\vec{I}$ the unit matrix. 
 
 In our simulation we use the following parameter values which are relevant for complex plasma experiments \cite{Couedel2011}:
 mass $m=8.0476 \times 10^{-13}~kg$,  damping rate $\nu=10~s^{-1}$, thermostat temperature $T=300~K$, screening length $\lambda=\Delta=300~\mu m$, particle charge $Q=8000~e$,
wake charge $q=-1600~e$ and wake-particle distance $\delta=90~\mu m$. These lead to an effective wake charge $\tilde{q}=0.2$, a dimensionless wake-particle distance $\tilde{\delta}=0.3$, a screening parameter $\kappa=1$ and a coupling parameter $\Gamma=Q^2/(k_BT\Delta)=1.187\times 10^{4}$. 
During the simulation time we slowly  decrease the confinement strength $C$. In particular, 
starting from a value $C_0=4 \times 10^{-9} N/m$ we decrease $C$ every 4 seconds by a small amount $\Delta C= 10^{-11} N/m$.

For the reverse simulation we start from  a square bilayer configuration for 
$C_0=2 \times 10^{-9} N/m$ and we increase the confinement strength by $\Delta C$
every 4 seconds.

\begin{center}
{ \textbf{ACKNOWLEDGEMENTS}}
\end{center}
This work was supported by the German Research Foundation (DFG) under Nos. LO 418/23-1 and IV 20/3-1. H. H. and C.-R. D. acknowledge the support by the
National Natural Science Foundation of China (NSFC), Grant No. 11975073. The authors would like to thank E. O\v{g}uz, V. Nosenko and H. Thomas for useful discussions.

\end{document}